\documentclass[aps,superscriptaddress,notitlepage,10pt,a4paper,twocolumn]{revtex4-1}

\usepackage[pdftex]{graphicx}
\usepackage{amsmath}
\usepackage{hyperref}
\usepackage{times}
\usepackage[centering,hmargin=2cm,vmargin=2.5cm]{geometry}

\newcommand{\OmegaN}{{\Omega_\textrm{N}}} 

\hypersetup
{%
pdftitle = {The phase shift induced by a single atom in free space},
pdfauthor = {Markus Sondermann},
colorlinks = {true},
urlcolor=blue,
linkcolor=blue
}

\begin{document}

\title{The phase shift induced by a single atom in free space}

\author{Markus Sondermann}
\email{markus.sondermann@fau.de}
\affiliation{Institute of Optics, Information and Photonics,
  University of Erlangen-Nuremberg, 91058 Erlangen, Germany}
\affiliation{Max Planck Institute for the Science of Light, 91058
  Erlangen, Germany}
\author{Gerd Leuchs}
\affiliation{Institute of Optics, Information and Photonics,
  University of Erlangen-Nuremberg, 91058 Erlangen, Germany}
\affiliation{Max Planck Institute for the Science of Light, 91058
  Erlangen, Germany}

\date{}

\begin{abstract}
In this article we theoretically study the phase shift a single atom
imprints onto a coherent state light beam in free space. 
The calculations are performed in a semiclassical framework.
The key parameters governing the interaction and thus the measurable
phase shift are the solid angle from which the light is focused onto the
atom and the overlap of the incident radiation with the atomic dipole
radiation pattern.
The analysis includes saturation effects and discusses the associated
Kerr-type non-linearity of a single atom.
\end{abstract}

\maketitle

\section{Introduction}

The interaction of light and single atoms in free space has received a
considerable amount of interest over the past years, see
Ref.~\cite{leuchs2013o} for an overview of recent achievements in this
field.
Here, the term \emph{free space} is used to describe a situation in
which the atom interacts with the whole continuum of the free-space
field modes and has also the characteristic free-space spatial
emission properties.
Besides other phenomena, the phase shift imprinted by a single
quantum system onto a coherent beam has been
studied in prior
experiments~\cite{aljunid2009,pototschnig2011,hetet2012}.
The reported phase shifts amount to  about 1$^\circ$ using a single
neutral atom~\cite{aljunid2009} and 3$^\circ$ for a single
molecule~\cite{pototschnig2011}.
Phase shifts of about 0.3$^\circ$ have been achieved recently for a
single ion~\cite{hetet2012}.
For the sake of simplicity, all kinds of quantum systems will be
denoted by the term 'atom' throughout this paper.

The maximum phase shift observed for a free-space setup is still an
order of magnitude below the values achieved with cavity quantum
electrodynamics setups~\cite{turchette1995,fushman2008,young2011}, but
a phase shift close to the maximum possible value of 180$^\circ$ has
not been observed in either system. 
Dispersive interaction has also been studied for an atomic ensemble
trapped in the evanescent field of a nano fibre~\cite{dawkins2011}.
However, the deduced phase shift per single atom does not exceed the
values measured so far in a free-space setup.

\begin{figure}
\centerline{\includegraphics{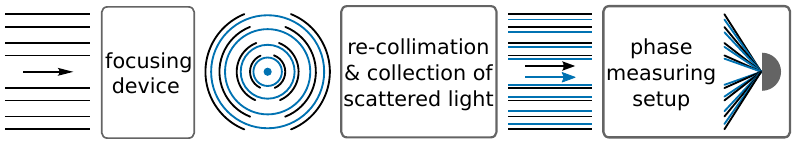}}
\caption{\label{fig:scheme}
Basic layout of a phase shift experiment with a single atom and a
coherent beam.
For further explanations see text.
}
\end{figure}

The typical phase-shift setup in free space can be simplified to the
scheme shown in Fig.~\ref{fig:scheme}.
The incident electromagnetic field mode is focused onto the atom by a
focusing device, e.g. a large numerical aperture 
lens~\cite{aljunid2009,pototschnig2011,hetet2012} or a parabolic
mirror~\cite{sondermann2007}.
Depending on the electric field strength acting upon the atom, 
the atom scatters a certain amount of dipole radiation which is
phase-shifted with respect to the incident field.
The phase of the scattered field is determined solely by the detuning
of the incident light from the atomic resonance.
The scattered radiation as well as the re-diverging incident field are
both collected by the same optical element.
This can be a second lens~\cite{aljunid2009,hetet2012}
or the focusing device itself.
The latter is the case e.g. in Ref.~\cite{pototschnig2011}, where a
reflective element retro-reflects the incident radiation towards the
focusing lens, or when using a deep parabolic mirror as envisaged
earlier~\cite{sondermann2007,lindlein2007}. 

After collection, scattered and incident radiation are processed in a
phase measuring setup.
This may be a Mach-Zehnder interferometer~\cite{aljunid2009}, a
heterodyning setup~\cite{pototschnig2011} or a scheme utilizing
polarization degrees of freedom~\cite{hetet2012}.
As part of the phase measuring setup the scattered and re-collimated
incident radiation are focused onto some detector where they interfere.
From this it is clear that the overall overlap of the scattered
radiation with the incident one plays a decisive role in determining
the measurable phase shift, see also Ref.~\cite{aljunid2009}.
Moreover, the amount of coherently scattered light determines the
impact of the phase of the scattered light onto the phase of the total
field, i.e. the superposition of scattered field and incident field.

Theoretical treatments can be found in several publications.
In Ref.~\cite{aljunid2009} the central parameter governing the phase
of the total field is the so called scattering ratio, i.e. the
ratio of scattered power to the incident power.
This parameter includes the overlap of the incident field with the
dipole field radiated by the atom.
The scattering ratio can reach a value of two when focusing from half
solid angle~\cite{zumofen2008,tey2009}, which is the upper limit
considered in Ref.~\cite{aljunid2009} leading to a maximum phase shift
approaching 90$^\circ$ for arbitrarily small but non-zero detunings.
This is in accordance with the findings of
Ref.~\cite{pototschnig2011} where the same maximum phase shift is
predicted. 
There, a single overlap parameter is used to account for the focusing
geometry as well as various aspects related to the use of a molecule.
Nevertheless, phase shifts of more than 90$^\circ$ are possible.
This is already evident from Eq.~26 of Ref.~\cite{zumofen2009}:
When considering a dipole wave incident from full solid angle, which
corresponds to a semi-aperture angle $\alpha=\pi$ in the nomenclature of
Ref.~\cite{zumofen2009}, one finds a phase shift of 180$^\circ$ on
resonance.
The same result is obtained when using the formulas of
Ref.~\cite{aljunid2009} and calculating the scattering ratio for
arbitrary solid angle, including the full solid angle case~\cite{leuchs2013o}.

Another aspect is the influence of saturation of the atomic transition
on the achievable phase shift.
This aspect has been treated in the calculations of
Ref.~\cite{pototschnig2011}, where also a corresponding measurement
has been performed, as well as by van Enk~\cite{vanenk2004}.
Although the latter paper is not explicitly devoted to the phase
shift problem, the results reported here and elsewhere can be
obtained by calculating the argument of Eq. 26 of
Ref.~\cite{vanenk2004}, using Eq. 41 of the same reference.
There, the treatment is fully quantum-mechanical.
However, also Ref.~\cite{vanenk2004} quantifies the similarity of the
incident wave with a dipole mode by using a single parameter, which is
inconvenient when modelling real experiments. 
The nonlinear phase shift induced onto a pulse containing more than
one photon has been derived in Ref.~\cite{koshino2008}, where
perfect coupling of atom and light field has been assumed implicitly.
Consequently, the influence of the key parameters discussed here has
not been treated. 

In this paper, we present a discussion of the phase shift induced by a
single atom in free space accounting for all of the above effects
in an explicit way.
Section~\ref{deriv} comprises the derivation of a formula for the
phase shift, followed by the treatment of some examples in
Sec.~\ref{examples}.
Finally, in Sec.~\ref{kerr} the inclusion of saturation effects is
used to derive a formula for the phase shift that is reminiscent of
the Kerr effect found in other non-linear optical media.

\section{Derivation of the phase shift}
\label{deriv}

We consider a two-level atom with upper level $|a\rangle$ and lower level
$|b\rangle$. 
It is located at the origin of the coordinate system. 
The atom is illuminated by a weak classical field of frequency
$\omega$ and amplitude $E_0$ at the place of the atom.
We take $E_0$ to be real.
Thus, the relative phase of the field at the place of the atom is zero.
Furthermore, $E_0$ is the amplitude of the field component parallel to
the atomic dipole.

The expectation value of the positive frequency part
of the electric field which is scattered by the atom is
given by \cite{scully1997} 
\begin{equation}
\langle\hat{E}^+(r,t)\rangle= \frac{\omega_0^2\mu \sin\vartheta}{4\pi\epsilon_0 c^2
  r}
\cdot \langle\hat{\sigma}_- (t-r/c)\rangle
\end{equation}
where $\omega_0$ and $\mu$ are the atomic transition frequency and the
dipole matrix element (taken to be real), respectively, $\vartheta$ is
the angle between the quantization axis and the point
$\vec{r}=r\cdot\vec{e}_r$ and $\hat{\sigma}_-$ is the atomic lowering
operator. 
The expectation value of the lowering operator is given by \cite{scully1997}
$\langle\hat{\sigma}_- (t)\rangle=\rho_{ab}(t)$
with $\rho_{ab}$ being the density matrix element describing the
polarization of the atom.

In the steady state, we have 
\begin{equation}
\rho_{ab}= -i \frac{\Omega_\textrm{R}}{2}\cdot (2\rho_{aa}-1)
\cdot \frac{\Gamma/2+i\Delta}{\Delta^2+\Gamma^2/4}
\end{equation}
with the spontaneous emission rate
$\Gamma=\omega_0^3\mu^2/(3\pi\epsilon_0\hbar c^3)$,
the Rabi frequency $\Omega_\textrm{R}=E_0\mu/\hbar$,
the detuning $\Delta=\omega-\omega_0$
and the density matrix element $\rho_{aa}$ gives the probability to
find the atom in the upper state. 
The steady state solution of the latter is given by
\begin{equation}
\rho_{aa}=\frac{\Omega_\textrm{R}^2}{4\Delta^2+\Gamma^2+2\Omega_\textrm{R}^2}
\end{equation}
which leads to 
\begin{equation}
\rho_{ab}=\frac{\Omega_\textrm{R}\cdot(i\Gamma-2\Delta)}
{4\Delta^2+\Gamma^2+2\Omega_\textrm{R}^2} \, .
\end{equation}
Thus, the scattered field amplitude is 
\begin{equation}
\label{eq:Esc}
E_\textrm{sc}=\frac{\omega_0^2\mu \sin\vartheta}{4\pi\epsilon_0 c^2 r} 
\cdot \frac{\Omega_\textrm{R}\cdot(i\Gamma-2\Delta)}
{4\Delta^2+\Gamma^2+2\Omega_\textrm{R}^2} \, .
\end{equation}
The phase of the scattered field is hence given by 
\begin{equation}
\varphi_\textrm{sc}=\arctan\left(-\frac{\Gamma}{2\Delta}\right)
=\arctan\left(\frac{2\Delta}{\Gamma}\right)
+\frac{\pi}{2} \,.
\end{equation}

For a wave of power $P$ incident onto the atom the field amplitude
parallel to the atomic dipole is given by \cite{sondermann2008}
\begin{equation}
E_0=\frac{\sqrt{2P}}{\lambda\sqrt{\epsilon_0 c}}
\cdot\sqrt{\Omega}\cdot\eta 
\end{equation}
with $\lambda=2\pi c/\omega$.
$\Omega$ is the effective solid angle over which the incident field
extends calculated weighting by the atomic dipole characteristics. 
It has a maximum value of $8\pi/3$. 
$\eta$ is the overlap of the incident field with the field emitted by
the atomic dipole, calculated only in the region covered by the
incident light.
Next, we insert this expression for $E_0$ into the definition of the
Rabi frequency, approximating $\omega\simeq\omega_0$.
Plugging the result into Eq.~\ref{eq:Esc} and integrating over the
full solid angle we arrive at the scattered power
\begin{equation}
P_\mathrm{sc}= P\cdot\frac{3}{2\pi}\cdot\Omega\eta^2\cdot
\frac{4\Delta^2/\Gamma^2+1}
{(4\Delta^2/\Gamma^2+1+\frac{3P\Omega\eta^2}{\pi\hbar\omega_0\Gamma})^2}
 \, . 
\end{equation}
The last term in the sum of the the denominator is the saturation
parameter on resonance
$s_0=2\Omega_\textrm{R}^2/\Gamma^2=3P\Omega\eta^2/(\pi\hbar\omega_0\Gamma)$. 

For the sake of simplicity we normalize the solid angle $\Omega$ to its maximum
value, $\OmegaN=\Omega/(8\pi/3)$, which results in
\mbox{$s_0=8P\OmegaN\eta^2/(\hbar\omega_0\Gamma)$}. 
Furthermore, the saturation parameter at non-zero detuning is
\mbox{$s=s_0/(1+4\Delta^2/\Gamma^2)$}.
This leads to
\begin{equation}
\label{eq:Psc}
P_\mathrm{sc}= \frac{4P\cdot\OmegaN\eta^2}
{(4\Delta^2/\Gamma^2+1)(1+s)^2}
 \, . 
\end{equation}

Now, we have all ingredients at hand to calculate the phase of the
field resulting from the superposition of the incident field and the
scattered field. 
However, one has to distinguish two scenarios. 
The first one is a symmetric setup in the sense that the re-diverging
incident field is collected with optics covering the same amount of
solid angle as for focusing.
Furthermore, the spatial radiation pattern of the incident light
after re-collimation is identical to the one before focusing.
This scenario is the one occurring in
Refs.~\cite{aljunid2009,pototschnig2011,hetet2012}.
In an asymmetric setup, the optics used for collection/re-collimation may cover a
different fraction of the solid angle than the one used for focusing,
as also treated e.g. in Ref.~\cite{zumofen2008} for the extinction of
a coherent beam.
Another example for an asymmetric setup is the usage of different types
of optics for focusing and collection, respectively~\cite{streed2012,jechow2013}.
Also every finite size parabolic mirror constitutes an asymmetric
setup as outlined in more detail below.

\subsection{Symmetric case}

First, we have to account for the fact that only the part of the
scattered field emitted into the solid angle cone of the transmitted
incident field has to be considered.
The power of this fraction is 
\begin{equation}
\label{eq:PscatOm}
P_\OmegaN=P_\textrm{sc}\cdot\OmegaN=
\frac{4P\cdot\OmegaN^2\eta^2}{(4\Delta^2/\Gamma^2+1)(1+s)^2}
 \, .
\end{equation}
The amplitude of the corresponding field mode is $A_\OmegaN\sim \sqrt{P_\OmegaN}$.
Taking the phase of the scattered field into account yields
$E_\OmegaN=A_\OmegaN\cdot e^{i\varphi_\textrm{sc}}$.
Furthermore, we have $E_\textrm{in}=A_\textrm{in}\sim\sqrt{P}$.
We also have to account for the Gouy phase shift of $\pi/2$ that the
transmitted wave experiences while re-diverging from the location of
the atom~\cite{tyc2012,zumofen2008,pototschnig2011,aljunid2009}.
We do this by $\varphi_\textrm{sc}\rightarrow
\arctan\left(\frac{2\Delta}{\Gamma}\right) +\pi$.  
Furthermore, only the part of the scattered wave that overlaps with the
transmitted incident wave can interfere with it.
Moreover, only the coherent part of the scattered light will
interfere with the incident light~\cite{wrigge2008}.
The coherently scattered power fraction is given by $1/(1+s)$~\cite{meschede2005}.
We take this into account by writing
\begin{equation}
E_\OmegaN\rightarrow \mathcal{E}_\OmegaN=E_\OmegaN\cdot
\eta \cdot (1+s)^{-1/2}\, . 
\end{equation}

The phase of the coherent superposition of incident and
scattered field can then be expressed as 
$\phi= \arg(\frac{E_\textrm{in}+\mathcal{E}_\OmegaN}{E_\textrm{in}})$
\cite{aljunid2009}. 
This leads to 
\begin{equation}
\phi=\arg\left(
1 + \frac{2\OmegaN\eta^2}{(1+s)^{3/2}\sqrt{1+4\Delta^2/\Gamma^2}}
\cdot \textrm{e}^{i\varphi_\textrm{sc}}
\right)\,.
\end{equation}
With $\sin\varphi_\textrm{sc}=-2\Delta/\Gamma/\sqrt{4\Delta^2/\Gamma^2+1}$
and $\cos\varphi_\textrm{sc}=-1/\sqrt{4\Delta^2/\Gamma^2+1}$
we finally arrive at
\begin{align}
\label{eq:phi}
\phi=\arg[&
(1+s)^{3/2}(1+4\Delta^2/\Gamma^2)-2\OmegaN\eta^2 \\
& - i\cdot 4\OmegaN\eta^2\Delta/\Gamma
] \, .\nonumber
\end{align}

\subsection{Asymmetric case}

We now treat the asymmetric case.
Using optics of different aperture for focusing and re-collimation
results in an dipole weighted solid angle covered by the
re-collimation optics $\OmegaN'\ne\OmegaN$. 
Thus Eq.~\ref{eq:PscatOm} changes to 
\begin{equation}
P_{\OmegaN'}=P_\textrm{sc}\cdot\OmegaN'=
\frac{4P\cdot\OmegaN\OmegaN'\eta^2}{(4\Delta^2/\Gamma^2+1)(1+s)^2}
\end{equation}
and we write $E_{\OmegaN'}=A_{\OmegaN'}\cdot e^{i\varphi_\textrm{sc}}$ with
$A_{\OmegaN'}\sim \sqrt{P_{\OmegaN'}}$.
A further consequence of such a scenario is that the overlap
parameter $\eta$, calculated only for the part of the solid
angle used for focusing, may change for the re-collimation optics.
Hence, for the part of the scattered light interfering with the
incident light we have
\begin{equation}
\mathcal{E}_{\OmegaN'}=E_{\OmegaN'}\cdot
\eta' \cdot (1+s)^{-1/2}\, . 
\end{equation}
Last but not least the case $\OmegaN'<\OmegaN$ induces a power loss to
the re-collimated incident beam. 
We account for this by introducing the parameter $p$ with $0\le
p\le1$.
When calculating the $\arg()$-function, we now write 
$\phi=\arg(\frac{\sqrt{p}E_\textrm{in}+\mathcal{E}_{\OmegaN'}}{\sqrt{p}E_\textrm{in}})$. 

With these modifications we arrive at the phase shift for the
asymmetric, i.e. general case:
\begin{align}
\label{eq:phiasym}
\phi=\arg[&
\sqrt{p}(1+s)^{3/2}(1+4\Delta^2/\Gamma^2)\\
& -2\sqrt{\OmegaN\OmegaN'}\eta\eta'
 - i\cdot 4\sqrt{\OmegaN\OmegaN'}\eta\eta'\Delta/\Gamma
] \,.\nonumber
\end{align}

\section{Examples}
\label{examples}

\begin{figure}
\centerline{
\includegraphics{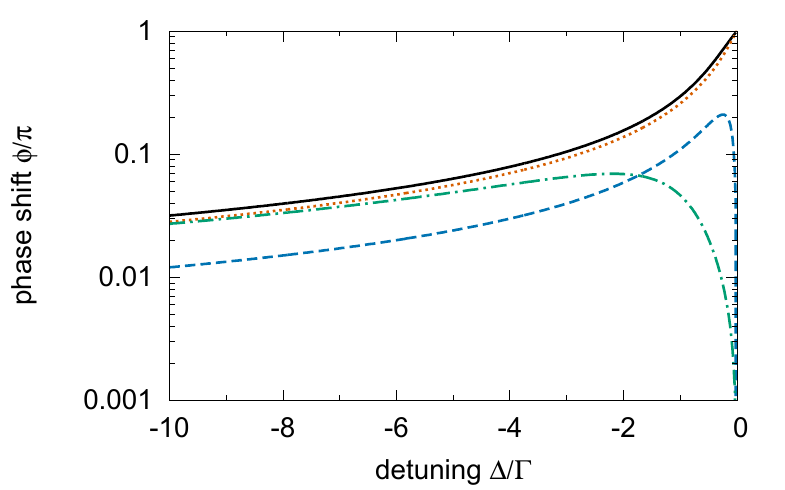}
}
\caption{\label{fig:examples}
Example calculations of the phase shift induced by a single atom.
Solid line: $\OmegaN=\eta=1$, $s_0=0$; 
dashed line: same but $\OmegaN=0.38$;
dotted line: $\OmegaN=0.94$, $\eta=0.98$, $\OmegaN'=0.88$, $\eta'=0.99$,
$p=0.97$, $s_0=0.1$; 
dash-dotted line: same but $s_0=10$.
For a description of the corresponding scenarios see text.
}
\end{figure}

The results of some example calculations are given in
Fig.~\ref{fig:examples}, starting the discussion with the
symmetric case.
Since the phase shift just changes sign when the detuning does, only
values for $\Delta/\Gamma\le0$ are plotted.
The solid line depicts the case that results in the maximum possible
phase shift at any detuning: focusing from full solid angle with a
dipolar radiation pattern and a negligible saturation parameter.

The dashed line depicts the case when focusing from 38\% of the solid
angle weighted with the radiation pattern of a linear dipole oriented
perpendicular to the optical axis.
The solid angle fraction corresponds to the one covered by a microscope
objective with a numerical aperture of NA=0.95.
Again, we assume perfect mode overlap and negligible saturation.

Next, we treat an example for the asymmetric case.
The dotted line corresponds to the experimental setup described in
Refs.~\cite{sondermann2007,golla2012} and a low but
non-negligible saturation parameter.
The setup consists of a parabolic mirror covering almost
the entire solid angle, described by a parameter $\OmegaN=0.94$.
As incident light field a radially polarized doughnut mode is
considered.
Such a mode is expected to couple efficiently to an atom with a linear
dipole transition located at the mirror's focus.
A doughnut mode with an overlap parameter of $\eta=0.98$ has been
achieved recently in experiments~\cite{golla2012}.
The dash-dotted line shows the phase shift for the same setup and
strong saturation of the atom.

However, using a parabolic mirror brings some intricate details that
necessitate the use of Eq.~\ref{eq:phiasym}.
Any ray propagating along the optical axis of the parabola which
enters the mirror at a distance $d$ to the optical axis leaves the
mirror at a distance $d'=4f^2/d$ with $f$ the focal length of the
parabolic mirror~\cite{leuchs2008}.
This has the following consequences:
For a finite parabolic mirror, the change $d\rightarrow d'$ entails
that rays entering the mirror close to the optical 
axis do not hit the parabolic surface a second time and are not
re-collimated.
This results in  a smaller effective solid angle for the re-collimated
incident light $\OmegaN'<\OmegaN$ and a re-collimated power-fraction
$p<1$.
Furthermore, a parabolic mirror reshapes the radiation pattern of the
incident beam. 
That is, after re-collimation by the parabolic mirror the transmitted
incident beam has another overlap $\eta'\ne\eta$ with the dipole mode
than upon focusing onto the ion.
In the present example, we account for all these effects by setting 
$\OmegaN'=0.88$, $\eta'=0.99$, and $p=0.97$.

Performing the calculations for the same setup using Eq.~\ref{eq:phi}
for the symmetric case, i.e. setting $\OmegaN'=\OmegaN$, $\eta'=\eta$
and $p=1$, leads to larger phase shifts. 
But the deviations are so small, on the order of 1.5\%, that
Eq.~\ref{eq:phi} will be used in the remainder of this paper for the
sake of simplicity. 
Nevertheless, one can construct realistic examples in which the
deviation is more pronounced.

\begin{figure}
\centerline{
\includegraphics{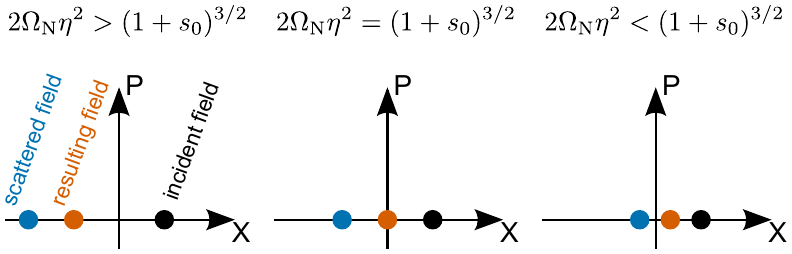}
}
\caption{\label{phase_space}
Illustration of the phase shift on resonance ($\Delta=0$) in a phase
space picture. 
}
\end{figure}

In what follows, we examine the phase shift close to resonance in more
detail.
For arbitrary $\eta$ and $\OmegaN$ the phase shift on resonance
is determined by the sign of the real part of the argument of the
$\arg$-function in Eq.~\ref{eq:phi}:
\begin{equation}
\label{eq:resonance}
\varphi_{\Delta=0} = \left\{
\begin{array}{c l}
\pi&\textrm{, if}\, 2\OmegaN\eta^2 > (1+s_0)^{3/2}  \\
0&\textrm{, else}
\end{array}
\right. 
\end{equation}
In other words, the solid angle fraction must be larger than
$(1+s_0)^{3/2}/(2\eta^2)$ for observing a non-zero phase shift on
resonance: 
Illumination has to occur from more than half
the solid angle (see also Ref.~\cite{leuchs2013o}).
However, even for $\OmegaN>1/2$ a low $\eta$ or a large saturation
parameter may result in zero phase shift.
Figure~\ref{phase_space} illustrates this discussion in a phase
space picture. 

\begin{figure}
\centerline{
\includegraphics{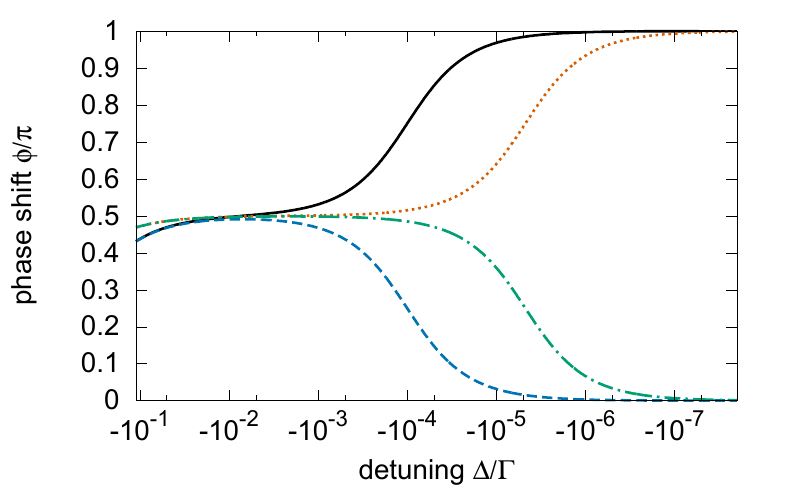}
}
\caption{\label{deltazero}
Phase shift close to zero detuning. Solid line: $s_0=0$, $\eta=1$,
$\OmegaN=0.5+10^{-4}$; 
dashed line: same but $\OmegaN=0.5-10^{-4}$; 
dotted line: $\OmegaN=\eta=1$, $s_0=\sqrt[3]{4}-1 - 10^{-5}$; 
dash-dotted line: same but $s_0=\sqrt[3]{4}-1 + 10^{-5}$.
}
\end{figure}

The change of the phase shift at $\Delta=0$ from $\pi$ to zero occurs
in an abrupt manner, i.e. the phase shift depends nonlinearly on the parameter
combination $\OmegaN\eta^2/(1+s_0)^{3/2}$.
This is illustrated in Fig.~\ref{deltazero} in more detail for
detunings $|\Delta|\ll\Gamma$.
The first example (solid line and dashed line) with the change of
$\OmegaN$ from slightly below to slightly above half solid angle may
seem unrealistic. 
However, this case is realizable by using a parabolic mirror covering
almost the entire solid angle and restricting the solid angle cone of
the incident light to a fraction of the mirror surface corresponding
to the values given above.

We conclude this section noting that it is obvious from
Eqs.~\ref{eq:resonance} and Fig.~\ref{phase_space} that 
the overall phase shift is not determined by the atom but rather by
the properties of the incident field and the set-up.
No matter how large the coupling efficiency is in dependence of $\eta$
and $\OmegaN$~\cite{leuchs2013o,golla2012}, the phase of the light
scattered coherently by the atom is fixed for a given detuning. 
This phase lag has been examined recently by a background
subtraction technique~\cite{jechow2013} in an experimental regime of
low coupling efficiency, i.e. exciting the atom from small solid
angle. 
The better the coupling efficiency and the lower
the saturation parameter, the larger is the amount of light scattered
coherently into the mode of the incident field.
This amount determines the phase of the superposition of incident and
scattered field.

\section{Intensity dependent phase shift}
\label{kerr}

The examples presented in Figs.~\ref{fig:examples} and~\ref{deltazero}
already demonstrated that the phase shift induced by a single atom is
strongly influenced by the intensity of the light driving the atom.
In what follows, the phase shift derived above is put into a shape
reminiscent of the typical formulas used to describe the optical Kerr effect.
One way of describing this effect is to write the
refractive index as~\cite{maker1964,chiao1964,Boyd1992}
\begin{equation}
\label{eq:standardkerr}
n=n_0+n_2\cdot I \, ,
\end{equation}
where $n_0$ is the weak-field refractive index and $n_2$ describes the
change of refractive index due to the intensity $I$ of the optical
field.
Corresponding more detailed expressions can be found for an
ensemble of two-level atoms in free space e.g. in Ref.~\cite{Boyd1992}
and e.g. in Ref.~\cite{hilico1992} for an an atomic ensemble in a
cavity.
To arrive at a similar expression in the scenario treated here requires
several approximations.

In order to have an interaction that is predominantly dispersive, we
assume a saturation parameter $s\ll 1$.
In other words, the atom is not excited on average and the light is
scattered coherently by the atom.
The condition of low saturation is met easiest at large detunings.
Therefore, we assume $|\Delta|\gg\Gamma$.
This includes the case $|\Delta|\ge\Gamma/2$ for which one finds that 
$(1+s)^{3/2}(1+4\Delta^2/\Gamma^2)-2\OmegaN\eta^2\ge0$ and hence
$|\varphi|\le\pi/2\ \forall\ \OmegaN, \eta, s$. 
In other words, we can rewrite Eq.~\ref{eq:phi} as
\begin{equation}
\varphi=\arctan\left(-
\frac{4\OmegaN\eta^2\Delta/\Gamma}{(1+s)^{3/2}(1+4\Delta^2/\Gamma^2)-2\OmegaN\eta^2}
\right)\, . 
\end{equation}
Furthermore, the large detuning approximation allows to 
replace the $\arctan$ function by its argument yielding
\begin{equation}
\varphi\approx
-\frac{4\OmegaN\eta^2\Delta/\Gamma}{(1+s)^{3/2}(1+4\Delta^2/\Gamma^2)-2\OmegaN\eta^2}
\, . 
\end{equation}
A Taylor expansion of the above equation around $s=0$ finally yields a
Kerr-type expression reading
\begin{eqnarray}
\label{eq:kerr}
\varphi& = & \varphi_0 -\frac{3}{2}\varphi_0\cdot s\, ,\\
\varphi_0& = &-\frac{4\OmegaN\eta^2\Delta/\Gamma}{1+4\Delta^2/\Gamma^2-2\OmegaN\eta^2}
\,.\nonumber
\end{eqnarray}

\begin{figure}
\centerline{
\includegraphics{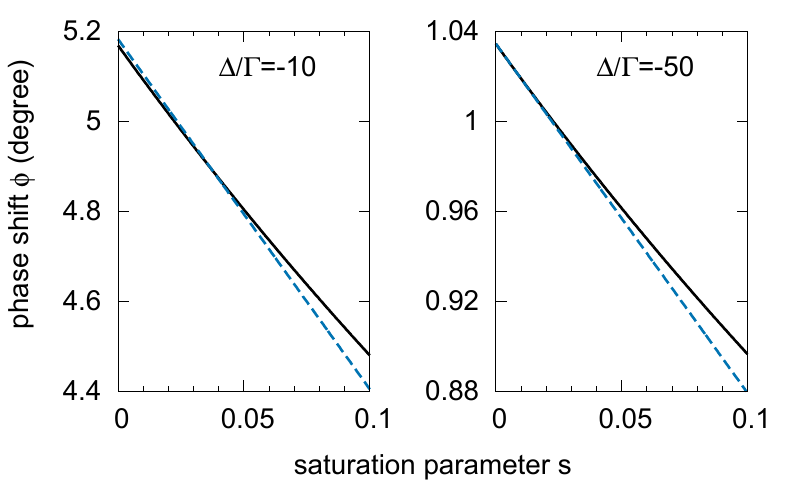}
}
\caption{\label{kerrfig}
Illustration of the single atom Kerr effect for 
$\OmegaN=0.94$, $\eta=0.98$, $\Delta/\Gamma=-10$ (left) and
$\Delta/\Gamma=-50$ (right). 
The phase shift is calculated using the full model via  Eq.~\ref{eq:phi} (solid lines)
and via Eq.~\ref{eq:kerr} approximating the atom as a pure Kerr-type
medium (dashed lines).}
\end{figure}

Figure~\ref{kerrfig} compares the above result to the full model of
Eq.~\ref{eq:phi}.
As to be expected, the quality of the approximate expression of
Eq.~\ref{eq:kerr} improves with increasing detuning and decreasing
saturation parameters.

\section{Concluding remarks}

As outlined above, the phase shift induced on the exciting field by a
single atom in free space is maximized by coupling the incident light
to the atom from full solid angle.
In this case, the phase shifts observed for large detunings are still
of considerable magnitude.
As evident from Fig.~\ref{kerrfig}, the phase shift observed at 50
linewidths detuning is on the order of the ones reported for low
detuning in previous experiments using free space setups.
This suggests that a single atom in free space might be a good
candidate for the realization of a quantum repeater scheme based on
dispersive light-matter interaction.
Such a scheme has been proposed by van Loock et al. for cavity
based setups~\cite{vanloock2006}.
When realizing such a system one has to balance all parameters carefully.
For example, the induced phase shift should exceed the uncertainty
of the phase of the incident coherent state, which is given by the
inverse of the square root of the state's amplitude if the latter is
sufficiently large.
Thus, on might be tempted to improve the performance by using
coherent states of larger amplitude.
But this in turn results in a larger saturation parameter and a
reduction of the imprinted phase shift, unless the temporal
width of the incident pulse is increased as well to maintain
constant incident power.
On the other hand, the pulse duration affects the success rate of
the repeater scheme. 
This brief discussion highlights that a detailed assessment of all
parameters using a free-space setup is desirable.
However, this is beyond the scope of the present paper and subject of
future work.

\begin{acknowledgments}
The authors acknowledge M. Bader, R. Maiwald and M. Fischer for useful
comments on the manuscript.
\end{acknowledgments}



\end{document}